\documentclass[twocolumn,preprintnumbers,amsmath,amssymb]{revtex4}

\usepackage{graphicx}
\usepackage{dcolumn}
\usepackage{bm}

\begin{document}


\title{Gallium Phosphide Photonic Crystal Nanocavities in the Visible}

\author{Kelley Rivoire}
\author{Andrei Faraon}
\author{Jelena Vuckovic}%
 \email{krivoire@stanford.edu}
\affiliation{%
E. L. Ginzton Laboratory, Stanford University, Stanford, CA 94305-4085
}%


\begin{abstract}
Photonic crystal nanocavities at visible wavelengths are fabricated in a high refractive index (n$>$3.2) gallium phosphide membrane.
The cavities are probed via a cross-polarized reflectivity measurement and show resonances at wavelengths as low as 645 nm at room temperature, with quality factors
between 500 and 1700 for modes with volumes 0.7$(\lambda/n)^3$. These structures could be employed for submicron scale optoelectronic devices in the visible, and for coupling  to novel emitters with resonances in the visible such as nitrogen vacancy centers, and bio- and organic molecules.

\end{abstract}

\maketitle

Photonic crystal nanocavities confine light in ultrasmall volumes, making them ideal for low-power, on-chip optoelectronic devices\cite{altug1, fushman1, tanabe}
as well for exploring fundamental light-matter interactions\cite{vuckovic1, noda1}.
Most experimental work in photonic crystal cavities has been done at telecom bands\cite{noda1}. This has been facilitated by well established fabrication procedures for semiconductors such as silicon and gallium arsenide, which have band gaps in the infrared. Photonic crystal devices operating in the visible part of the spectrum, however, could serve as light sources and spectroscopic devices operating below 750 nm. Photonic crystal cavities with resonances in the visible could also be coupled to novel light emitters such as nitrogen vacancy centers\cite{lukin, gruber}, fluorescent molecules\cite{moerner}, and visible colloidal quantum dots\cite{moerner2}.

To date, most photonic crystal devices operating in the visible have been fabricated in one of three material systems: gallium nitride, silicon nitride, and AlGaInP. While devices with wavelengths as short as blue can be fabricated in gallium nitride\cite{hu1,lee}, fabrication processes in this system are difficult, and the low refractive index of gallium nitride (n=2.4) limits the device quality factors\cite{lidzey}. Low refractive index is also the main drawback of the silicon nitride system\cite{benson,makarova}. Two groups have fabricated AlGaInP-based photonic crystals in the red\cite{fitzgerald, scherer} with InGaP quantum wells. This system, however, requires precise control over the fraction of each material in the quaternary system to obtain the proper band gap and lattice constant, and is lattice matched to a GaAs substrate, which is absorbing in the visible.

In this letter, we show that photonic crystal devices at visible wavelengths can be fabricated in gallium phosphide. Gallium phosphide is a high refractive index (n=3.25 at 700 nm, n=3.44 at 555 nm at room temperature\cite{nelson}) III-V semiconductor with an indirect band gap at 555 nm at room temperature. Although gallium phosphide is not typically used for devices requiring very high brightness because of its indirect band gap, incorporating InP quantum dots or quantum wells \cite{hatami}, or colloidal quantum dots should greatly increase the quantum efficiency. Additionally, gallium phosphide has a relatively low surface recombination rate, estimated to be roughly an order of magnitude less than that of gallium arsenide\cite{baca}, because of its high phosphorus content, which has a passivating effect\cite{wang}. Surface recombination is particularly important for photonic crystal devices because of the additional exposed surface area from etched air holes.

\begin{figure}[h]
\includegraphics[width=8.5cm]{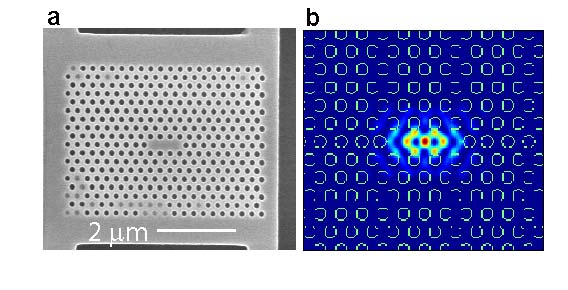}
\caption{\label{fig:pc2}(a) Scanning electron microscope image of a fabricated photonic crystal membrane after undercut of sacrificial layer with lattice constant a=246 nm, hole radius, r/a=0.32, slab thickness d/a=0.57. (b) Finite difference time domain simulation of electric field intensity inside the cavity for the high Q mode.}
\end{figure}

Linear three hole defect (L3)\cite{akahane} cavities were first simulated and then fabricated in a 140 nm thick GaP membrane (Fig. 1). Simulation results indicate that quality factors above 10,000 can be obtained for this type of cavity (three holes removed, holes next to cavity displaced outward by $0.15a$) using a triangular lattice of lattice constant $a$ with slab thickness $d/a$=0.55 and hole radius $r/a$=0.3 with no other design modifications. The electric field profile for the fundamental L3 cavity mode is shown in Fig. 1b.

To fabricate these structures, the 140 nm thick GaP membrane was grown on the top of a 1 $\mu$m thick sacrificial AlGaP layer. Patterns were first defined in ZEP520 photoresist by electron-beam lithography and then transferred into the GaP membrane by a chlorine-based reactive ion etch. Excess photoresist was removed with oxygen plasma, and the sacrificial layer was undercut with hydrofluoric acid to yield suspended membrane structures with high index contrast. Scanning electron microscope images of one of the fabricated membranes can be seen in Fig. 1a and 1c.

\begin{figure}[h]
\includegraphics{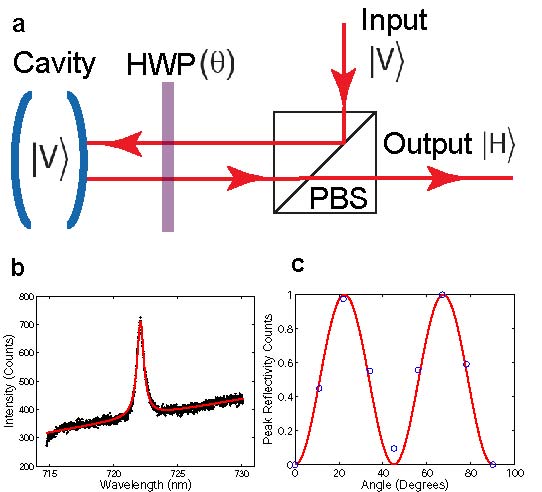}
\caption{\label{fig:pc} (a) Experimental setup. The V-polarized cavity is probed in a cross-polarized setup using a PBS and a HWP. The cavity signal observed at the output follows a $\text{sin}(4\theta)$ dependence where theta is the HWP setting ($\theta$=0 corresponds to V polarization).
(b) Spectrum of a cavity measured with configuration of Fig. 2a with HWP at $\theta=67^{\circ}$. The cavity resonance is at 722 nm at room temperature. A fit to a Lorentzian (solid line) gives Q=1100. (d) (Circles) Normalized counts in cavity peak as a function of HWP angle (background subtracted).  (Line) Fit with $\sin(4\theta)$.}
\end{figure}

The fabricated photonic crystal resonators were probed using a confocal cross-polarized reflectivity measurement\cite{altug2, englund_nature} technique as depicted in Fig. 2a.  This technique uses polarization control to achieve a high signal-to-noise ratio and allows probing of cavities with no internal light source. A vertically polarized (V) probe is directed through a polarizing beam splitter (PBS) and a half wave plate (HWP) onto the photonic crystal cavity, which has a mode that is also vertically polarized (V). The reflected output is observed through the PBS, which acts as a horizontal (H) polarizer. The primary polarization of the cavity mode is vertical (V). When the angle between the HWP fast axis and the vertical direction ($\theta$) is set to zero, the cavity coupled light is reflected with V-polarization and does not transmit through the beam splitter. Rotation of the HWP allows part of the cavity-coupled light to be transmitted at the PBS into the output port with intensity following a $\text{sin}(4\theta)$ dependence (Fig. 2(b,c)). A tungsten halogen lamp source was used as a wide broadband signal at the input. The output field was measured using a spectrometer with a liquid nitrogen cooled CCD.

\begin{figure}[h]
\includegraphics[width=8.5cm]{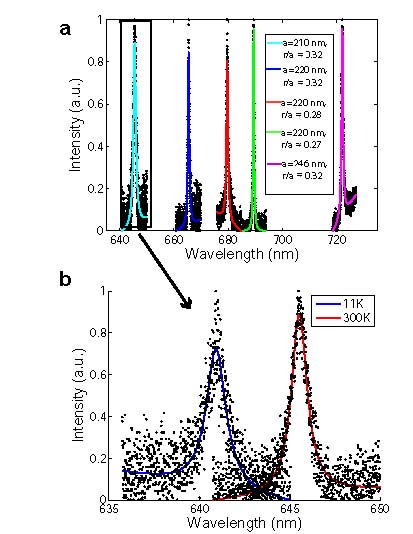}
\caption{\label{fig:pc2} (a) Shift of photonic
 crystal resonance at room temperature as a and r/a are changed. Solid lines are fits to Lorentzians. (b) Cavity resonance measured at room temperature (blue) and low temperature (red). Solid lines indicate fits to Lorentzians.}
\end{figure}

To fabricate cavities with resonant wavelengths spanning a large interval, lattice constant a (210nm$<$a$<$246nm) and hole radius (0.25$<$r/a$<$0.35) were modified. These changes in the fabrication parameters resulted in cavities with resonance wavelength spanning from 645 nm to 750 nm as shown in Fig. 3. (Plots in Fig. 3 are shown with background subtracted.) The measured resonances have shorter wavelengths than predicted from finite difference time domain simulations (by about 0.05$a/\lambda$). We believe this difference is a result of imperfect selectivity between the sacrificial layer and membrane during the wet etch step, which reduces the membrane thickness. Quality factors measured experimentally are between 500 and 1700. This is about one order of magnitude smaller than predicted by simulations because of imperfections introduced in the fabrication process.

The shortest resonant cavity wavelength observed at room temperature was 645 nm with quality factor 610. The resonance wavelength depends on the temperature-dependent refractive index of the membrane. To shift the cavity resonance to lower wavelengths, the sample was cooled to 11K in a continuous flow liquid helium cryostat. The change in temperature caused the resonance to shift from 645 nm to 641 nm (in the red wavelenth range) and the quality factor to decrease to 490 as shown in Fig. 3. With improved fabrication, devices operating close the 555 nm band gap of gallium phosphide could be made; such devices could be used as green LEDs and lasers. Quality factors could also be improved by using modified designs with simulated Q values more than a order of magnitude higher\cite{englund}. However, these Qs can still enable a significant Purcell effect, up to 185, as mode volume is only 0.7$(\lambda/n)^3$.

In summary, we demonstrate a new gallium phosphide-based materials system for photonic crystals in the visible. We probe linear three-hole defect cavities in reflectivity and observe resonances from 645 nm to 750 nm at room temperature with quality factors as high as 1700. To our knowledge, these are the shortest wavelength photonic crystal cavities fabricated in a high-index material.

We expect this materials system to allow access to a new range of photonic crystal devices in the visible, including LEDs and lasers. Additionally, the gallium phosphide system will provide opportunities to couple photonic crystal cavities to emitters in the visible such as colloidal quantum dots, nitrogen vacancy centers,
biomolecules, or organic molecules.

Financial support was provided by the MARCO Interconnect Focus Center, CIS Seed Funding, and the NSF graduate fellowship.  This work was performed in part at the Stanford Nanofabrication Facility of NNIN supported by the National Science Foundation under Grant No. ECS-9731293.

\end{document}